\documentclass{article}

\usepackage[english]{babel}

\usepackage[letterpaper,top=2cm,bottom=2cm,left=3cm,right=3cm,marginparwidth=1.75cm]{geometry}

\usepackage[version=3]{mhchem}
\usepackage{amsmath}
\usepackage{graphicx}
\usepackage{chemformula}
\usepackage[colorlinks=true, allcolors=blue]{hyperref}
\usepackage{multirow}
\usepackage[authoryear,compress]{natbib} 

\title{Evolution of the Archean Atmosphere}
\author{Colin Goldblatt, Jake K. Eager-Nash, and Julia E. Horne \\
School of Earth and Ocean Sciences, University of Victoria. czg@uvic.ca}

\begin{document}
\maketitle

\begin{abstract}
Archean atmospheric evolution is the transition from an abiological atmosphere, to an atmosphere for which the composition and therefore climate is highly altered by life. We review the key processes and transitions in this evolution. 
\end{abstract}

\section{Introduction}

The Archean is the Eon of greatest transition in the Earth system, manifest in the Atmosphere. Biological processes are a dominant control of the modern atmospheric composition; this was well established by the end-Archean, when biogenic oxygen transformed atmospheric chemistry. Yet the initial conditions for the Archean atmosphere were those of the lifeless Hadean, dominated by the high energy processes of planetary accretion. The transition from an abiological atmosphere to an atmosphere dominated by life took place over a number of evolutionary steps, the process of which are the substantive topic of this Chapter. 

The Archean Eon's boundaries are officially set at arbitrary dates (Geological Standard Stratographic Ages of 4.03 -- 2.5 Ga), but for a process oriented review it is more meaningful to view this in terms of actual Earth system events. Our Eon's name derives from the Greek $\alpha\phi\chi\eta$, \emph{in the beginning}, intended as the beginning of life \citep{Nisbet1991}. While the date of this is yet to be constrained and perhaps never will be, it is the logical beginning of this narrative. The end sits naturally at the Great Oxidation of Earth's atmosphere, when accumulation of waste biogenic oxygen transformed Earth's chemistry. We refer to most times by era: Eoarchean (4.03 -- 3.6 Ga), Paleoarchean (3.6 -- 3.2 Ga), Mesoarchean (3.2 -- 2.8 Ga), and Neoarchean (2.8 -- 2.5 Ga).

This short Chapter gives one of many of the narratives of the Archean atmosphere that can be told. A lucid account of the processes of Atmospheric Evolution is in a classic textbook by Jim \citet{Walker1977}, one of the forefathers of this field. In a recent textbook, \citet{CatlingKasting2017} offer an exhaustively detailed graduate-level treatment. \citet{Goldblatt2018} gives a brief review of the processes of atmospheric evolution. \citet{CatlingZahnle2020} provide a recent comprehensive review of the Archean atmosphere, synthesising the records and contemporary debates (we do not seek to replicate that, but suggest it be read alongside this contribution). We stand on all these shoulders.

This Chapter focuses on the key processes and transitions which lead to the evolution of the Archean atmosphere. We first describe the controls on climate, then on atmospheric composition, both in an Archean context. We then describe some of the major atmospheric transformations that occurred, usually as a result of biological innovation. 

\section{Controls on Climate}\label{climate}

Atmospheric composition is a primary determinant of climate. The control of climate is thus a key lens through which we describe Archean atmospheric evolution. 

A robust result of stellar evolution models is that the Sun, as with all comparable stars, has brightened through its main sequence lifetime. Consequently, the energy received by Earth has increased 30\% over time; as such, insolation was approximately 75\% of modern in the Eoarchean and 80\% in the Neoarchean. There is abundant evidence of ordinary fluvial sediments in the Archean but only occasional glaciation \citep{Nisbet1987}; the question ``what kept early Earth warm'' is known as the \emph{Faint Young Sun Problem} (FYSP; \citealp{Sagan1972,feulner12,Charnay2020}), which frames our discussion.

There are a few primary controls on global mean surface temperature. Stellar output and distance from the star determine the incident solar radiation at the top of the atmosphere, and fraction $1-\alpha_p$ is absorbed, where planetary albedo $\alpha_p$ is the fraction of sunlight reflected. Reflection occurs via scattering by molecules (so depends on the atmospheric pressure), by cloud particles or other aerosols, and by the surface (the ocean reflects least, snow and ice most, and land surfaces an intermediate amount). Some sunlight is absorbed in the atmosphere (about 25\% today, mostly by water vapour), and certain aerosols can also absorb (minor today). Planets are warmed by a greenhouse effect, when thermal radiation is absorbed and re-emitted in the atmosphere, by gasses or clouds; the strength broadly depends on the temperature difference between the atmosphere (cold aloft) and the surface, given that colder bodies radiate less. Greenhouse gas absorption depends on both the logarithm of gas abundance, and on pressure broadening by ambient air \citep{Pierrehumbert2011}, making composition and pressure both relevant. High clouds are effective greenhouse agents, as they are cold. We address each of greenhouse gases, pressure, clouds, and haze, later in this section. 

The equator to pole temperature gradient is the second most important property of the climate. Earth is heated by sunlight primarily in the low latitude, so polar warmth depends on both heat transport and the greenhouse effect. With less insolation but a stronger greenhouse, the Archean would have had a weaker equator to pole temperature gradient \citep{Goldblatt2021}. This would make glaciation harder to initiate but more likely to extend to lower latitudes if it did. 

\subsection{Greenhouse gases} \label{ghgs}
Carbon dioxide (\ce{CO2}) and water vapour (\ce{H2O}) are Earth's primary greenhouse gasses; both absorb well in the infrared, but at different wavelengths. Water is unique in that its atmospheric concentration is controlled physically: given the abundant ocean, atmospheric water depends directly on saturation vapour pressure, which depends exponentially on temperature. Thus water abundance is a climate feedback, but cannot keep Earth warm by itself. 
Carbon dioxide corresponds to the redox state of the surface Earth system, and is photochemically stable, so given Earth's large carbon inventory, \ch{CO2} can accumulate in the atmosphere (we discuss the geochemistry in section \ref{inorgc}), and is the go-to gas for primarily resolving the FYSP.

Methane (\ce{CH4}) is important as a secondary greenhouse gas. Absorbing at different wavelengths, it is complementary to \ch{CO2} and water vapour, but inherently weaker; its reputation as a stronger greenhouse gas arises from its present low concentration, so a smaller number of moles emitted will change the logarithm of concentration more \citep{Byrne2014a}. It is a ubiquitous biogenic carbon product (see section \ref{orgc}), but is subject to photochemical destruction. 

At high concentrations of methane, which are invoked for the Archean \citep{CatlingZahnle2020}, its climate effects are more complex. At higher methane to carbon dioxide ratios, methane's warming potential diminishes as it absorbs sunlight and warms the stratosphere \citep{Byrne2015,Eager-Nash2023}, while ethane (\ce{C2H6}) may be photochemically produced and complement methane as a greenhouse gas\citep{haqq-misra08}. However, if the methane to carbon dioxide ratio increases beyond 0.1, a photochemical haze will form \citep{Zahnle1986,trainer06}, complicating matters (Sec. haze).  

Numerous other greenhouse gases have been proposed, including ammonia (\ce{NH3}; \citealp{Sagan1972}), nitrous oxide (\ce{N2O}; \citealp{Buick2007}), carbonyl sulfide (\ce{COS}, \citealp{Ueno2009}; see \citealp{Byrne2014b} for a summary and comparison), but none have survived the ravages of further study. Ammonia is the best greenhouse gas, absorbing well where water does not \citep{Sagan1972, Byrne2014b}, but has the disadvantages of being photochemically unstable \citep{Kuhn1979,Kasting1982}, very soluble \citep{Levine1980}, and subject to immediate biological uptake \citep{Horne2024}. Nitrous oxide is equivalent in inherent strength to \ch{CO2} \citep{Byrne2014a}, but rapidly photolysed \citep{Roberson2011}.

\subsection{Atmospheric pressure} \label{atmsize}
Atmospheric pressure, which is directly related to the mass of Earth's atmosphere, has two contrasting effects on mean surface temperature. Pressure broadening of greenhouse gas absorption lines strengthens the greenhouse effect and warms. Conversely, Rayleigh scattering increases albedo and cools. At the expected pressure range for the Archean, from a fraction of a bar to a couple of bars \citep{Goldblatt2009,CatlingZahnle2020}, pressure broadening dominates and gives 2--5\,K warming per doubling \citep{Goldblatt2009,WolfToon2014}. The transition to Rayleigh dominance happens at 2--5\,bars (higher when there are more greenhouse gases) \citep{Goldblatt2017}. If the lowest constraints on past pressure \citep{Som2016} are to be believed, then a big fraction of the Archean atmosphere would have needed to have been \ch{CO2} in order to keep Earth warm. 

Pressure  also affects the equator-to-pole temperature gradient: a more massive atmosphere could transport more heat and weaken the gradient, whereas a lower pressure would give a bigger temperature gradient \citep{KaspiShowman2015}. 

\subsection{Clouds} \label{clouds}
Despite early investigations \citep{Rossow1982}, clouds were long neglected in FYSP work. One-dimensional radiative--convective models (RCMs) dominated the field until a decade ago, as General Circulation Models (GCMs) were too expensive for deep palaeoclimate (with some exceptions: \citealp{Jenkins1993}), and no-one really knew how to parameterise cloud feedbacks in 1-D. Indeed, a leading modeller omitted clouds entirely in his RCM, adjusting surface albedo upward to compensate for the net cooling that clouds give in the modern \citep{Kasting1984}, and that assumption stuck for a generation. Using 1-D parameter exploration, \citet{GoldblattZahnle2011cloud} showed that realistic changes to clouds could make a substantive contribution to warming the early Earth. A new wave of GCM application to the FYSP found that less \ch{CO2} was required to maintain temperate climate than previously thought \citep{WolfToon2013,Charnay2013}. This is because the amount of low cloud decreases when the solar constant is lower but the greenhouse effect stronger \citep{Goldblatt2021}. \citet{Charnay2020} review the impact of GCMs in resolving the FYSP.

\subsection{Photochemical haze}\label{haze}
Aerosols produced photochemically have likely been important too. The classic case is a hydrocarbon haze forming when \ch{CH4}/\ch{CO2}$\geq$0.1.  Climate calculations for the Archean using the optical properties of anoxic Titan-like hazes indicate that a thick haze (corresponding to \ch{CH4}/\ch{CO2}$>$0.1) would absorb sunlight high in the atmosphere and cool the surface \citep{haqq-misra08,Arney2016,Mak2023}, whereas a thin haze (corresponding to \ch{CH4}/\ch{CO2}$=$0.1) could lead to a warming due to haze interaction with clouds and water vapour \citep{Mak2023}. 

However, more recent optical calculations of hazes made  with some oxygen in the precursor gas, as there would have been on Archean Earth, indicate hazes that absorb little sunlight, while still scattering \cite{Ugelow2018}. Thus the climate effect is a little hazy.  

\subsection{Summary: Archean climate} \label{climatesum}
In summary and in general, a temperate Archean climate was maintained by a stronger greenhouse, with a secondary contribution from cloud feedbacks. However, as the Archean spans 1.5 billion years, it is reasonable to think of vast climate changes occurring; by Phanerozoic (538.8\,Ma -- present) comparison it took a mere 50 million years to transition from the early Eocene (Ypresian, 56.0--47.8\,Ma) hothouse to today's glacial climate. The last generation of climate models showed that the Archean could be kept warm; the next generation will describe the physics of diverse Archean climates.  
 
\section{Atmospheric composition: nested cycles and the bio-geo atmosphere} \label{atmcomp}

Atmospheric composition is largely determined by geochemical cycling, and both are profoundly influenced by biology. We use the term ``bio-geo'' atmosphere to distinguish from the abiological atmospheres of the other planets. 

Thus, our process-oriented description of the Archean atmospheric evolution necessarily involves description of how the major geochemical cycles affect the atmosphere. These cycles are nested; some processes are fast and affect climate on human timescales, yet others are so slow that they only operate over Eons (a small rate multiplied by a billion years is generally a big flux!). On the scale of atmospheric evolution, all of the cycles interact. Thus, understanding all of the timescales and all of the cycles is necessary in the study of the Archean atmosphere. 


\subsection{Carbon} \label{ccycle}

Carbon dioxide is the dominant greenhouse gas in determining Earth's temperature, yet the atmospheric \ch{CO2} inventory is a tiny fraction of the total atmosphere-ocean inventory, which in turn is a minuscule fraction of the whole Earth inventory. Feedbacks on \ch{CO2} level are complicated, stabilizing or destabilizing climate on different timescales \citep{Archer2010}. Carbon is the fundamental building block of life, so life is deeply entwined in the carbon cycle. Moreover, links between carbon and oxygen cycling have played a critical role in \ch{O2} evolution. The carbon cycle is thus at the heart of atmospheric evolution. 

The carbon cycle can be split into two branches, inorganic (concerning oxidised carbon only), and organic (coupling oxidised and reduced carbon, primarily through life), with the former having the most critical role in climate regulation.

\subsubsection{Inorganic Carbon} \label{inorgc}

Most of Earth's carbon in inorganic, with key reservoirs being atmospheric \ch{CO2} ($9 \times 10^{14}$\,kg\,C), ocean Dissolved Inorganic Carbon (DIC, $4 \times 10^{16}$\,kg\,C), carbonate sediments ($2.5 \times 10^{15}$\,kg\,C) and carbonate rocks ($6 \times 10^{19}$\,kg\,C), and total mantle carbon is around $7 \times 10^{20}$\,kg\,C \citep{Goldblatt2018}. Understanding and reconstructing atmospheric \ch{CO2} thus requires understanding how Earth's carbon inventory is split both between the atmosphere and ocean, and between atmosphere--ocean and solid Earth.

Atmospheric \ch{CO2} readily dissolves in the ocean, where it hydrates to form carbonic acid (\ce{H2CO3}), which may then dissociate to bicarbonate and carbonate ions (\ce{HCO3-} and \ce{CO3^{2-}}, respectively; \citealp{ZeebeWolfGladrow}):  
\begin{eqnarray}
\ce{CO2} + \ce{H2O} &\ce{<=>} & \ce{H2CO3}    \nonumber \\
\ce{H2CO3} & \ce{<=>} & \ce{H^+} + \ce{HCO3^-}  \label{e-carbpartition}\\
\ce{HCO3^-} & \ce{<=>} & \ce{H^+} + \ce{CO3^{2-}}  \nonumber
\end{eqnarray}
The direct consequence of this is a large ocean store of inorganic carbon; only \ce{CO2}, typically the smallest of the reservoirs which comprise DIC ($\text{DIC} =  [\ch{CO2}] + [\ce{HCO3^-}] +  [\ce{CO3^{2-}}]$), exchanges with the atmosphere. 

This system requires specifying \emph{two} parameters to solve. In addition to DIC, the second natural controlling parameter is alkalinity, a complicated concept which may be understood as either the capacity to buffer against acid addition until a pH of 4 \citep{EmersonHamme2022}, or the net negative charge in the inorganic carbon system \citep{ZeebeWolfGladrow}. In the simplified case of Eq. \ref{e-carbpartition}, $\text{Alk} = [\ce{HCO3^-}] +  2[\ce{CO3^{2-}}]$. A higher ratio of DIC to alkalinity gives a larger fraction of atmosphere--ocean carbon as \ch{CO2}, and conversely a relative increase in alkalinity would decrease \ch{CO2}. While DIC and alkalinity are the most natural controls, specifying any two parameters are sufficient. 

Calcium carbonate precipitation
\begin{equation}
\ce{Ca^{2+} + CO3^{2-} <=> CaCO3} \label{e-calcite}
\end{equation}
and deposition into sedimentary rock removes DIC and alkalinity. Kinetic inhibition means that the ocean must be supersaturated with respect to the calcium carbonate for this to occur. Today this is mostly performed by calcifying organisms, buffering supersaturation to the minimum level at which these precipitate \ch{CaCO3}. Carbonate precipitation reduces supersaturation, decreases pH, and increases the atmospheric fraction of atmosphere--ocean inorganic carbon, because 2 moles of alkalinity are removed for every one mole of DIC in Eq. \ref{e-calcite}. Significant innovations in biological calcification occurred at the Cambrian explosion and in the Mesozoic \citep{Ridgwell2005margeo}. By implication, calcite supersaturation in the Archean was significantly higher than it is today. 

Table \ref{t-aocarb} shows the split between atmosphere and ocean carbon for possible Neoarchean conditions. The Neoarchean atmospheric fraction is higher than in the modern, but across the expected supersaturation range $10<\Omega<20$, this varies by 50\%. Thus, the scope for a reduction in saturation state to rapidly increase atmospheric \ch{CO2} is considerable. 

\begin{table}
    \centering
    \begin{tabular}{|l|c|c|c|c|}
        \hline
         &  Pre-ind. & \multicolumn{3}{|c|}{Neoarchean} \\
        \hline
        Calcite saturation, $\Omega$ & 4.65 & 5 & 10 & 20\\
        \hline    
        DIC ($\mu$mol/kg) & 2,000 &20,380 & 28,480 & 40,060  \\
        Alkalinity ($\mu$mol/kg)& 2,280 &19,480 & 27,800 & 39,810  \\
        pCO2 ($\mu$bar)&  & 30,000 & 30,000 & 30,000 \\
        pH & 8.17 & 7.17  & 7.32 & 7.47 \\
        Atmosphere C ($10^{16}$\,kg\,C)& 0.061 & 6.41 & 6.41 & 6.41\\
        Ocean C ($10^{16}$\,kg\,C)& 3.24 & 33.0 & 46.2 & 65.0\\
        Total C ($10^{16}$\,kg\,C)& 3.30 & 39.5 & 52.6 & 71.4\\
        Atmosphere fraction & 0.019 & 0.19 & 0.14 & 0.099\\
        \hline    
    \end{tabular}
    \caption{Possible states of the inorganic carbon system for the Neoarchean, compared to pre-industrial. For Neoarchean runs, p\ch{CO2} is set to 30,000\,ppmv, which would give a temperate climate \citep{Goldblatt2021}. Calculations are performed with pyCO2SYS \citep{Humphreys2022}}
    \label{t-aocarb}
\end{table}

The massive reservoir of carbonate rock has necessarily grown through Earth history. In evolutionary models, we have assumed that much of this growth happened in the Archean from an initial zero inventory \citep{Horne2024}. In essence, this means the processing of volcanic outgassing through the atmosphere, ocean, and into carbonates. As the nutrient P is deposited into carbonate rocks, and will be supplied again by the weathring of these, change in the carbonate reservoir also affects primary productivity \citep{Horne2024,Alcott2024}. 


\subsubsection{Climate Feedbacks} \label{climfeedbacks}
\citet{Archer2010} describes how different feedbacks in the inorganic carbon cycle, on different timescales, are stabilizing or destabilizing. On a hundred year timescale, modern anthropogenic \ch{CO2} emissions are buffered by ocean uptake in a stabilizing feedback. Yet on thousand to ten-thousand year timescales the glacial--interglacial record shows that warming increases atmospheric \ch{CO2}, a destabilizing feedback (this is linked to ocean circulation changes and changes in the atmospheric fraction of atmosphere--ocean carbon). On hundred-thousand year timescales, the silicate weathering feedback removes more atmosphere--ocean carbon to the solid Earth in warm and high-\ch{CO2} conditions, via enhanced chemical weathering, a long-term stabilizing feedback. 

Given the long duration of the Archean, it is common to only think in terms of the stabilizing silicate weathering feedback, but this is an error. The shorter timescale, destabilizing, processes also occur, and will facilitate rapid climate change at times. Moreover, the temperature towards which silicate weathering will regulate depends on both the volcanic carbon source and non-climate controls of weathering. For example, low volcanism could decrease the steady state temperature to below the threshold for runaway ice expansion: while generally stabilizing, silicate weathering feedback does not preclude carbon cycle changes causing glaciation. 

\subsubsection{Organic carbon} \label{orgc}

Key modern organic carbon reservoirs are living biomass ($6 \times 10^{14}$\,kg\,C), atmospheric \ch{CH4} ($1.7 \times 10^{12}$\,kg\,C), organic sediments ($1.6 \times 10^{15}$\,kg\,C) and buried organic carbon sedimentary rocks ($1.5 \times 10^{19}$\,kg\,C; \citealp{Goldblatt2018}). Modern ocean biomass is one-hundredth of the total, which is a more relevant reference value for the Archean, as trees dominate modern biomass. 

Biological productivity is carbon fixation: reducing \ch{CO2} to organic carbon. Oxygenic photosynthesis has dominated this flux since it evolved (see section \ref{oxyphoto}), and it can be reversed by aerobic respiration
\begin{equation}
\ce{CO2 + H2O <=>[\text{oxygenic photosynthesis}][\text{aerobic respiration}] CH2O + O2} \label{e-oxps}
\end{equation}
The modern rate of this is $2.5 \times 10^{14}$\,kg\,C\,yr$^{-1}$ \citep{Crockford2023}. Through Earth history, it is estimated that $\sim 10^{23}$ to $\sim 10^{24}$\,kg\,C have been processed through photosynthesis \citep{Crockford2023}, which is hundreds to thousands times greater than Earth's total carbon stock. Archean oxygenic primary productivity is estimated to have been one-hundredth to one-tenth of the modern rate, \citep{Crockford2023, Horne2024}, or $2.5 \times 10^{12}$ to $2.5 \times 10^{13}$\,kg\,C\,yr$^{-1}$. Taking a nominal age for the origin of oxygen producing photosynthesis as 3.5\,Ga, Archean processing of carbon would have been $2.5 \times 10^{21}$ to $2.5 \times 10^{22}$\,kg\,C, double to twenty times the whole Earth carbon inventory. That is to say, the carbon cycle is absolutely dominated by life today, and likely was in the Archean too. 

During the Archean, many of the environs where decomposition occurred were profoundly anoxic \citep{Lyons2014}, so a major decomposition pathway was fermentation followed by methanogenesis, net: 
\begin{equation}
\ce{CH2O -> 1/2CO2 + 1/2CH4} \label{e-methanogen}
\end{equation}
so the combined effect of oxygenic photosynthesis and decomposition is 
\begin{equation}
\ce{1/2CO2 + H2O -> O2 + 1/2CH4} \label{e-netpsmethanogen}
\end{equation}
producing methane and oxygen in stoichiometric balance. This is the major source of methane, which was an important greenhouse gas. This cycle is closed by atmospheric methane oxidation, and the consequences of this for oxygen are discussed in the following section.

\subsection{Oxygen} \label{oxygen}
The Archean surface was mildly reducing, with a Great Oxidation marking the end Archean and transition into the Proterozoic \citep{Lyons2014}. Much has been said, including in this Volume, on the evidence for this, so herein we focus entirely on the atmospheric processes involved.  


The paradox of Archean oxygen is that oxygenic photosynthesis evolved deep in the Archean, yet the Great Oxidation was substantively delayed. This is naturally explained via the chemistry of the atmosphere: two distinct levels of oxygen are permitted, separated by the development of the ozone layer \citep{Goldblatt2006, Gregory2021, Wogan2022, Garduno2023}. Archean oxygenesis would firstly have sustained oxygen levels in the parts per billion range, higher or lower depending on the prevailing sources or sinks of oxygen. At a threshold around 0.1 ppm, there was sufficient oxygen that an ozone layer formed via the \citet{Chapman1930} reactions. The ozone layer provided a UV shield which slowed atmospheric photo-oxidation reactions, which in turn caused oxygen levels to increase by a few orders of magnitude, to $\sim 0.1$\% \citep{Gregory2021, Wogan2022, Garduno2023}. This level gives oxidising conditions at the surface. Thus, from a purely atmospheric point of view, the Great Oxidation was the formation of the ozone layer \citep{Goldblatt2006}.

Through the Archean the balance of oxygen sources and sinks determined both the contemporaneous level of atmospheric oxygen, and the redox evolution. The most important flux to consider is the flux of oxygen to the atmosphere: oxygenic primary productivity less aerobic respiration (or decomposition via reaction with other oxidized species), which is in effect the fraction of primary productivity that is decomposed by methanogenesis, and less any consumption by methanotrophs. This given a flux of oxygen stoichiometrically balanced with methane (Eq. \ref{e-netpsmethanogen})
which is then closed via atmospheric oxidation \citep{Goldblatt2006}, net:
\begin{equation}
\ce{O2 + 1/2CH4 -> 1/2CO2 + H2O} \label{e-atmosox}
\end{equation}
The larger this input flux, the higher the oxygen level, so variations in primary productivity would have driven variations of oxygen levels. 

Opposing this is a net consumption of oxygen other than by the oxidation of biogenic methane, that is, consumption of what we term a net source of reductant. This includes oxidation of reduced volcanic or metamorphic gases \citep{Claire2006}, oxidative weathering of rocks if \ch{O2} levels are sufficiently high, and less excess organic carbon burial (burying reductant that would otherwise be subject to aerobic respiration or conversion to methane is a net oxygen source). Archean oxygen levels, and a threshold for the Great Oxidation, is then related to a ratio of this net oxygen production to net reductant flux \citep{Goldblatt2009o}.

Long term trends in Archean oxygen would have thus been caused by secular trends in primary production, changes in the oxidation state of the solid Earth, and growth of the buried organic carbon reservoir. 

The Archean is the emergence of planetary scale life; so by definition it began with negligible buried organic carbon, and may have ended with a substantial reservoir (our model estimates put this at a tenth of modern; \citealp{Horne2024}). In this interactive model of the carbon, oxygen, and nutrient cycles, the slowing of organic carbon accumulation (and associated oxygen flux) by nutrient limitation proved to be critical in delaying the Oxidation \citep{Horne2024}.

Long term planetary oxidation is thought to be dominantly caused by hydrogen escape to space, with the hydrogen derived from methane \citep{Catling2001,CatlingZahnle2020}. As diffusion limited hydrogen escape is directly proportional to hydrogen mixing ratio at the homopause \citep{hunten1976}, a more reduced atmosphere will, perhaps counter-intuitively, facilitate a faster oxidation of the Earth. A xenon isotope proxy of hydrogen escape demands abundant methane, greater than 5,000 ppm, at least for periods of time throughout the Archean \citep{Zahnle2019}. Redox balance models indicate that atmospheric steady state requires methane levels be proportional to the net reductant input less growth in the buried organic carbon reservoir \citep{Goldblatt2006,Goldblatt2009o}. This implies that a very substantive reductant flux from the solid Earth was required.

\subsection{Nitrogen}

Atmospheric nitrogen levels were for a long time generally considered to have been unchanging. However, more recent work \citep{Goldblatt2009} proposed that not only has the nitrogen level likely changed, but this would have consequences on climate. The present atmospheric inventory, 0.78\,bar, is $4.0\times10^{18}$\,kg\,N, whereas by contrast to carbon, the ocean reservoir is small $2.4\times10^{16}$\,kg\,N \citep{Johnson2015}.

There are large and geologically accessible nitrogen reservoirs in the crust and mantle which, critically, have been processed through the atmosphere \citep{Johnson2015}. Nitrogen gets into rock as ammonium (\ch{NH4+}) substituted for potassium ions (\ch{K+}), with the ammonium generally sourced from biological fixation of atmospheric nitrogen. 

Crustal nitrogen, estimated to be $2.7\times10^{18}$\,kg\,N, comprises two thirds as much as the atmosphere, and has increased over time \citep{Johnson2017}. Given the timescale of continental growth and the growth in N concentration, this reservoir unavoidably came from the atmosphere via nitrogen fixation. 

Mantle nitrogen reservoirs are best estimated from noble gas systematics. Nitrogen has similar mantle solubility to argon. Nitrogen is correlated with \ch{^{40}Ar}, the daughter product of \ch{^{40}K}, not to primordial \ch{^{36}Ar}. Therefore mantle nitrogen is of subducted origin \citep{Marty1995,Marty2012}, fixed from the atmosphere, and substituted for \ch{K+} in rocks. Estimates are $(7.2\pm5.9)\times10^{18}$\,kg\,N in a ``MORB-like'' source, plus $(17\pm15)\times10^{18}$\,kg\,N in a ``high-N'' source, total $(24.2\pm16)\times10^{18}$ \citep{Johnson2015}. That is, the mantle contains at least as much nitrogen as the atmosphere, and estimated six times as much, and this nitrogen has been assimilated into the mantle from the atmosphere. 

Determining a nitrogen history is challenging. There are only a handful of proxy constraint on past nitrogen, from 3.5\,Ga to 2.7\,Ga, which indicate similar atmospheric levels to today \citep{Som2012,Marty2013} or perhaps lower \citep{Som2016}, but there is no time series through Earth history. Estimates of nitrogen flux into and out of the mantle for the present day imply a residence time of many-billion years relative to mantle exchange \citep{Goldblatt2009}, which is difficult to reconcile with inventory data, but models indicate that fluxes could have been an order of magnitude higher in the past \citep{Johnson2018}.

An alternative proposal to reconcile low p\ch{N2} measurements relies on high-temperature abiological reduction of N and diffusion into the mantle early, in a magma ocean phase. Oxidation of the mantle would shift the balance of the stable nitrogen phase from \ch{NH4+}, which is soluble, to \ch{N2} which would outgas \citep{Wordsworth2016}. 

\section{Biological transitions and atmospheric consequences}

Having established the major processes controlling the Archean atmospheric evolution, we now describe some of the key transitions which caused change. Earth's modern atmosphere is in profound chemical disequilibrium due to life \citep{lovelock1972}, and we emphasise the evolutionary steps which lead life to generate this disequilibrium during the Archean.

\subsection{Chemolithoautotrophy: the origin} 

Molecular phylogeny tends towards the conclusion that the last universal common ancestor (LUCA) was a chemolithoautotroph \citep[e.g.][]{Weiss2016}. Chemolithoautotrophs acquire both carbon and energy from the environment. The energy must come from exploiting existing chemical disequilibrium, such as oxidising reduced volcanic gases, in order to fix carbon, and the ecosystem would have been \emph{energetically limited}. As a result, the primary productivity of a chemolithoautotrophic ecosystem is significantly lower than a phototrophic ecosystem \citep{Sauterey2020}. Moreover, an entirely chemolithoautotrophic ecosystem will \emph{reduce} the amount of disequilibrium. 

Methanogenesis using molecular hydrogen as a substrate is seen as an important metabolism, and ecological models indicate that it can support a high methane flux \citep{Kharecha2005,Sauterey2020,Eager-Nash2024}. 
In these weakly reducing atmospheres, a fraction of the methane is photochemically oxidized to \ch{CO}, which would then accumulate (a photochemical \ch{CO} runaway). It is thus thought that CO-consuming organisms evolved relatively early to utilise this energy gradient \citep{Ferry2005}, preventing \ch{CO} runaway and leading to a high ratio of methane to carbon monoxide \citep{Thompson2022}.



\subsection{Phototrophy: generating disequilibrium} \label{phototrophy}

Photoautotrophs capture photons, and use this energy to fix carbon, rather than relying on chemical gradients. This is more efficient, permitting a step increase in carbon fixation. This new energy source for life, in the form of low entropy solar photons, means that ecosystems can generate disequilibrium. This origin would have been when the Archean atmosphere began to deviate most rapidly from its abiological precursor. 

It is thought that photosynthesis was initially anoxygenic, using electron donors other than water (e.g. molecular hydrogen, hydrogen sulphide, iron), and not producing oxygen \citep[see a review by][]{Hohmann-Marriott2011}. Anoxygenic photosynthesis was, therefore, limited by the availability of these electron donors (e.g. molecular hydrogen, hydrogen sulphide, iron), and the ecosystem was \emph{substrate limited}.

Nevertheless, anoxygenic photosynthesis likely supported a step increase in carbon fixation rates. Consumption of this organic carbon by methanogens (Eq \ref{e-methanogen}) may well have produced a higher methane flux. Reservoirs of buried organic carbon, sufficient to influence atmospheric evolution, would have begun to accumulate. 


\subsection{Oxygenic photosynthesis: nutrient limitation}  \label{oxyphoto}

Oxygenic photosynthesis changed everything. Using ubiquitous water and carbon dioxide substrates, this eliminated limitation by electron donors and massively expanded productivity. In the modern context, marine primary productivity is limited by nutrient availability, primarily N and P, and there is every reason to expect this \emph{nutrient limitation} was the case from soon after the evolution of this metabolism. Models indicate the development of oxygenic photosynthesis drove an increase in primary productivity by two orders of magnitude over anoxygenic photosynthesis \citep{Ward2019, Horne2024}. 

The atmospheric consequences would have been profound. Most directly, the introduction of oxygen, the toxic waste gas from Eq \ref{e-oxps}, to parts per billion level. Next, much more rapid accumulation of the organic carbon reservoir, providing an additional long-term carbon sink and facilitating long-term oxygen accumulation. More generally though, the progress from substrate limitation expanded the capacity of life to influence every other part of Earth's geochemistry, and generate disequilibrium in the atmosphere.  

Oxygenic photosynthesis evidently evolved sometime during the Archean, but when remains subject to debate (which is explored elsewhere in this Volume), so we keep our discussion minimal. For example, interpretations of iron isotopes from Isua call for iron oxidation by free oxygen and an Eoarchean origin \citep{Czaja2012}. Even for a more sceptical view, abundant stromatolite records from 2.8\,Ga support an origin by the Neoarchean \citep{olsonPhotosynthesisArcheanEra2006}. Historically,  the apparent paradox of a later Great Oxidation was used as an argument against early origin of oxygenesis, but this is now resolved via atmospheric chemical mechanisms (sec \ref{oxygen}).

\subsection{Nitrogen fixation}

The evolution of biological nitrogen fixation, giving organisms the ability to break the triple bond of \ch{N2}, would have provided for a substantive increase in productivity. Abiotic nitrogen fixation is rather limited, relying on lightning \citep{Kasting1981b} or photochemical HCN formation \citep{Zahnle1986,Tian2011}. Thus the ability of organisms to fix their own nitrogen would have facilitated a step change in primary productivity: without it, oxygenic photosynthesis would have been profoundly limited by N. 

The ability to draw-down \ch{N2}, and ultimately sequester in rocks as \ch{NH4^+}, will have profoundly affected atmospheric \ch{N2} evolution. Without biological nitrogen fixation, atmospheric \ch{N2} could only have increased through time. With ample time for the mantle to outgas through Earth history, the consequence could have been a $\sim 3$\,bar nitrogen atmosphere today. 

More generally, nitrogen fixation would have facilitated rapid nitrogen cycling. Further atmospheric consequences of this would have been sources of ammonia and nitrous oxide. While, as noted above (section \ref{atmcomp}), no modelling group has succeeded in accumulating enough of these to cause much warming, their destruction by photolysis means that they necessarily affected the atmospheric chemistry. 

Evidence puts an origin of Nitrogen fixation as Mesoarchean, or even Eoarchean. Nitrogen isotopes from 3.3 -- 2.8 Ga provide evidence of cycling of biogenic N \citep{stuekenIsotopicEvidenceBiological2015}, and the nitrogen isotope record since 3.2\,Ga is incompatible with an lightning source \citep{Barth2023}. Molecular clocks suggest a 3.1--2.7Ga origin \citep{Parsons2021}. High nitrogen content in 3.8\,Ga metasediments is modelled as most likely of biological origin \citep{stuekenNitrogenAncientMud2016}.

Nitrogen fixation operates via the nitrogenase enzyme, and constraints on this may have influenced the atmospheric evolution. Nitrogenase requires a metal cofactor, either iron, vanadium, or molybdenum \citep{HayMele2023}. Iron was abundantly available in the reducing Archean ocean whereas molybdenum, which is sourced from oxidative weathering of the continents, would have been limited. It has been suggested that this Mo-limitation could have limited Archean nitrogen fixation, and by extension limited primary productivity \citep{zerkleMetalLimitationCyanobacterial2006}. Nitrogenase is inhibited by oxygen; modern nitrogen fixers must temporally or spatially separate nitrogen fixation from oxygen production to circumvent this \citep{boydNewInsightsEvolutionary2013}, and this might be a challenge in Archean oxygen oases too.

\subsection{Biogenic enhancement of calcium carbonate precipitation}


Biogenic enhancement of carbonate precipitation likely evolved by the Palaeoarchean. Carbonate stromatolites are found at 3.4 Ga \citep{Allwood2006} and perhaps earlier. \citet{Xiang2024} describe three different types of Palaeorchean carbonate ``factories'', on an abiotic to biotic, or hydrothermal to microbe-mediated, gradient. First, an ocean crust carbonate factory forms abiotic carbonate in pillow basalts. Second, an organo-carbonate factory features carbonate mineralized in association with organic molecules, either biotic or abiotic. Third, and critical for the argument here, a microbial carbonate factory includes carbonate precipitation controlled by microbial extracellular polymeric substances. 

Thus, biological innovation and changes in depositional setting will have affected Archean carbonate saturation, though are unlikely to have decreased this to Phanerozoic levels. Table \ref{t-aocarb} shows that such changes (e.g. in $10<\Omega<20$) could have rapidly affect Archean p\ch{CO2}. Likewise, geological forcings such as changes in shelf area, changing neritic carbonate deposition (e.g. lowering supersaturation when there is more shelf; \citealp{Ridgwell2005margeo}) would have applied throughout the Archean.



\subsection{A model of Atmospheric Evolution}

\begin{figure}[tp]
    \centering
    \includegraphics[width=\textwidth]{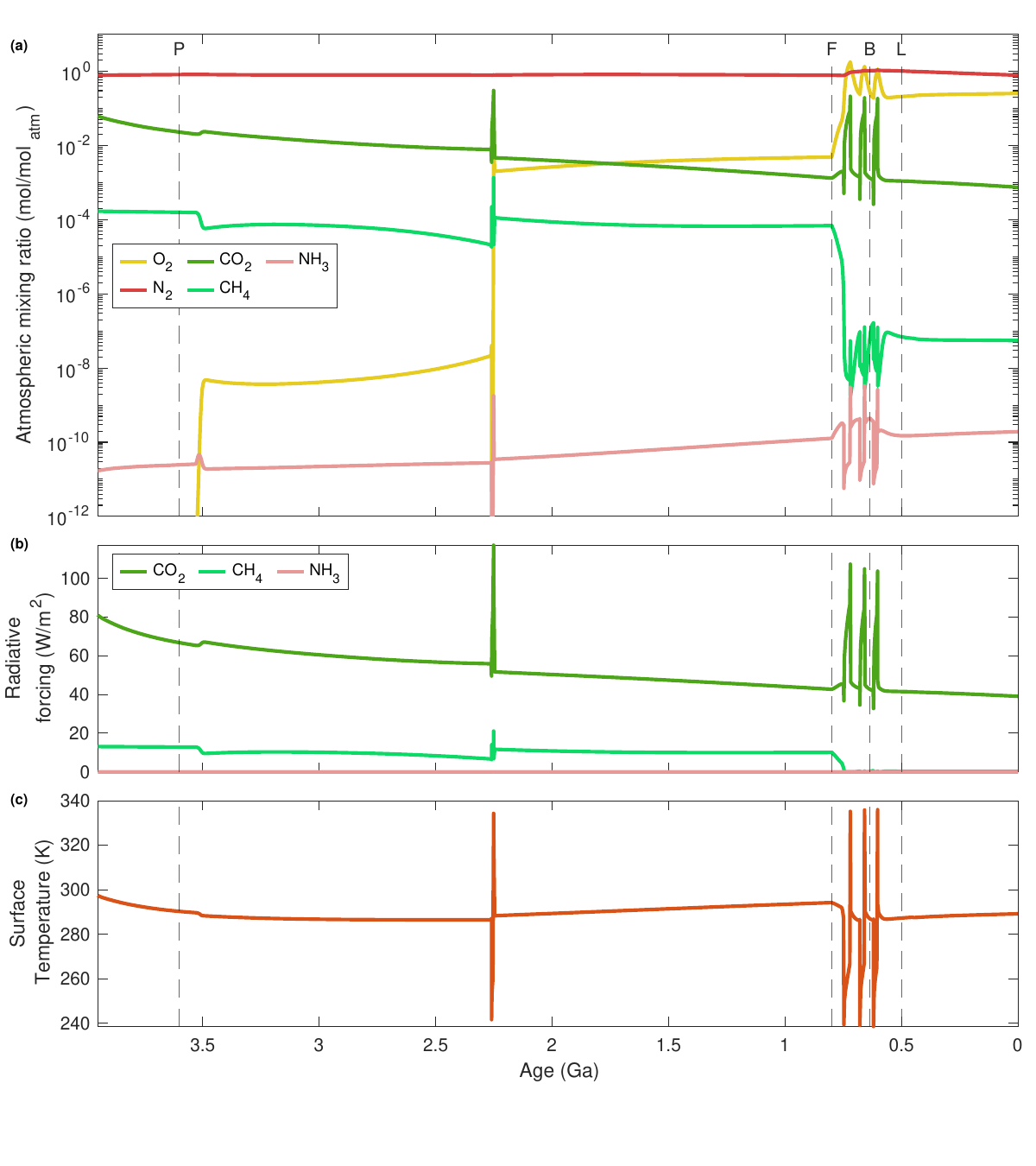}
    \vspace*{-15mm} 
    \caption[{EONS} Model of the Archean atmosphere]{\textbf{{EONS} Model of the Archean atmosphere} Atmospheric chemistry and climate output from \emph{EONS} using nominal tuning parameters and assumptions. (a) atmospheric composition, (b) radiative forcings from greenhouse gases, (c) surface temperature. Dashed vertical lines denote forced evolutionary transitions: \textit{P} for oxygenic photosynthesis, \textit{F} for fungi, \textit{B} for increased body sizes, and \textit{L} for vascular land plants.} \label{fig:atm}
\end{figure}

In Fig \ref{fig:atm}, we present model atmospheric evolution from our Earth's Oxygenation and Natural Systematics (EONS) model \citet{Horne2024}. This models the coupled dynamical evolution of the carbon, oxygen, nitrogen, and phosphorous cycles, from the Archean to the modern. It is the first model that we know of with this scope, and facilitates quantitative hypothesis testing for Archean atmospheric evolution, in a framework of coupled biogeochemical cycles. The model starts with anoxygenic photosynthesis and nitrogen fixation evolved, whereas the version shown here additionally includes a parameterized ice-albedo feedback; all other features are as \citet{Horne2024}. 

We discuss some features of the model, in relation to the preceding discussion. The nominal run represents a conservative ``best guess'' at an atmospheric evolution. While broadly consistent, we highlight a few interesting discrepancies as opportunities to improve model understanding of the Archean. 


Atmospheric carbon is dominated by \ce{CO2}, and this provides most of the radiative forcing to resolve the FYSP. Limited continental exposure slows alkalinity supply and carbonate sequestration (though changes in carbonate saturation are not implemented yet). 

Oxygen rises on the origin of oxygen producing photosynthesis, then remains $\sim 10$\,ppb until the Great Oxidation. The timing of the Great Oxidation is tuned via a quite conservative linear decline in net reductant input to the atmosphere-ocean through Earth history. 

Methane has a biological source and responds to oxygen: decreasing as oxygen levels increase due to rapid photo-oxidation. This is seen in the immediate drop following the initial oxygen rise, and renewed decline as oxygen levels increase prior to the Great Oxidation. This affects \ch{CO2} as well, because this is the product of methane oxidation.

Methane levels are overall lower than Xenon isotopes demand \citep{Zahnle2019}. Increasing the net input of reductant during the Archean would likely resolve this in the model, but this would need to be explained in a way that is consistent with mantle oxidation records. 

This nominal run does not give any substantial variation in \ce{N2} levels. However, our treatment of mantle outgassing and subduction zone processes are simplified, without the temperature and redox dependencies that other models \citep{Johnson2018} have implemented. That is to say, our model implies that major nitrogen variation would need to be geologically mediated. 

Though we track ammonia, and do see variations in this, they are never climatically relevant. High p\ce{CO2} keeps the ocean relatively acidic in the Archean, so most dissolved fixed nitrogen is \ce{NH4+} (which does not exchange with the atmosphere) rather than \ce{NH3}.  

A global glaciation occurs at the end of the Archean. Temperatures are close to the glaciation threshold, and the trigger is a decline in carbon dioxide and methane \emph{prior to} the Great Oxidation, and unstable carbon cycle feedbacks take over as temperature declines. The rise of oxygen does not occur until \emph{after} the glaciation, when the hot post-glacial climate enhances weathering fluxes and nutrient supply. While a single glaciation is shown here, superimposing changes in volcanism or subduction rates can give sequences of global glaciations (as seen in Fig \ref{fig:atm} for the Neoprotoerozoic). 

There is a small (2\,K) but rapid drop in temperature at the origin of oxygenic photosynthesis, driven by the decline in methane radiative forcing that is only partially offset by an increase in \ce{CO2}. In a different model tuning, with higher methane, there might be a bigger cooling. With a later origin or expansion of oxygenesis, this would be a potential cause of the Mesoarchean Pongola glaciation  \citep{youngEarthOldestReported1998,dewit5GaHydrothermalFields2016}.

\section{Conclusions }

In \emph{The Biosphere}, Vernadsky described the biosphere as the living world at the surface of the Earth, and the atmosphere as a non-living part of the biosphere. The atmosphere is of life, but not alive, and through controlling climate sets the environment for life.

Almost every atmospheric gas is controlled by a mix of biological and geological processes (the exceptions are water vapour, controlled by evaporation, and the noble gases which are aloof of chemistry). Biology affects \ch{N2}, \ch{O2}, \ch{CO2}, \ch{CH4}, \ch{N2O}, \ch{NH3}, \ch{CO}, and other gases. These biological influences all began in the Archean, and transformed the atmosphere. 

In this framework, Archean atmospheric evolution is the atmosphere becoming part of the biosphere.

\bibliographystyle{apalike}
\bibliography{Lib}

\end{document}